\documentclass[%
notitlepage,
superscriptaddress,
longbibliography,
 amsmath,amssymb,
 aps,
]{revtex4-2}

\usepackage{amsmath}
\usepackage{bm}
\usepackage{graphicx}
\usepackage{dcolumn}
\usepackage{bm}
\usepackage{hyperref}
\usepackage{epstopdf}
\usepackage[mathlines]{lineno}
\usepackage{multirow}    
\usepackage{booktabs}    
\usepackage{float}
\usepackage{subcaption}
\usepackage{xcolor}
\makeatletter
\newcommand{\doi@}[1]{\href{https://doi.org/#1}{#1}}
\DeclareRobustCommand{\doi}{\hyper@normalise\doi@}
\makeatother
\begin{document}
\title{Exploring Neural Network Surrogates for High-Order Mesh-Free Interpolants}

\author{Lucas Gerken Starepravo}
\email{lucas.gerkenstarepravo@postgrad.manchester.ac.uk}
\affiliation{School of Engineering, The University of Manchester, Manchester, UK}

\author{Georgios Fourtakas}
\affiliation{School of Engineering, The University of Manchester, Manchester, UK}

\author{Steven Lind}
\affiliation{School of Engineering, Cardiff University, Cardiff, UK}

\author{Ajay Harish}
\affiliation{School of Engineering, The University of Manchester, Manchester, UK}

\author{Jack R. C. King}
\affiliation{School of Engineering, The University of Manchester, Manchester, UK}

\begin{abstract}
\vspace{0.5cm}\noindent
Mesh-free numerical methods offer flexibility in the discretisation of complex geometries, showing significant potential for problems where mesh-based methods struggle. Although high-order approximations can be obtained through consistency-correction linear systems, such approaches remain prohibitively expensive for Lagrangian formulations, which commonly exhibit low-order convergence. Here we investigate the use of machine learning (ML) to bridge this gap, developing two strategies to couple multilayer perceptrons (MLPs) with the Local Anisotropic Basis Function Method (LABFM) as an exemplar high-order mesh-free method. In the first strategy, neural networks are trained to directly surrogate the high-order kernel; in the second, surrogate models are developed to compute the solution of the dense, low-rank linear systems arising in high-order mesh-free discretisations. The first strategy yields qualitative agreement with validation data but only marginally outperforms inconsistent smoothed particle hydrodynamics (SPH) kernels, with divergent behaviour observed for the Laplacian operator. The second strategy produces solution vectors with mean absolute errors of $\mathcal{O}(10^{-4}$--$10^{-5})$, replicating LABFM second-order convergence rate up to a resolution-dependent limiting accuracy and achieving up to a $5\times$ speedup at equivalent accuracy. However, sensitivity analyses reveal that higher-order approximations impose increasingly stringent accuracy requirements on the predicted solution vector, representing a fundamental challenge for current neural network architectures.

\end{abstract}

\maketitle

\section{Introduction}
In the context of the numerical solution of partial differential equations (PDEs), mesh-free methods are those that require no information on connections between collocation points, allowing for collocation points to be distributed in a disordered/unstructured manner through the computational domain. This is in contrast to mesh-based methods, where the topological connections between discrete computation nodes are used in calculations. These meshless collocation points can be placed to satisfy specific criteria, such as resolution requirements and geometric boundaries, allowing efficient handling of complex geometries and dynamic problems, addressing limitations commonly encountered in traditional mesh-based methods.

Developed in the 1970s and originally formulated for astrophysical problems, smoothed particle hydrodynamics (SPH) has been adapted to a broad range of physical systems and is today perhaps the most widely used mesh-free numerical method~\cite{gingold_smoothed_1977,lucy_numerical_1977}. In this framework, the continuum domain is represented by a set of (typically) Lagrangian particles, with interactions governed by a smoothing kernel. In its standard formulation, SPH exhibited low-order convergence~\cite{Liu2005}. This limitation is characteristic of many Lagrangian mesh-free methods~\cite{Lind_review}, which tend to operate at low-order accuracy since high-order approximations often require solution of linear systems. Although high-order approximations can be advantageous---achieving improved accuracy at the same resolution compared to low-order approximations---the computational cost becomes prohibitive in Lagrangian frameworks due to the need for frequent solution of linear systems, as the Lagrangian nodes move in time. 

Nearly 50 years after the introduction of SPH, a plethora of mesh-free methods has emerged. The reproducing kernel particle method (RKPM)~\cite{RKPM_liu1995}, generalised finite difference method (GFDM)~\cite{jensen_finite_1972, benito_solving_2007}, the generalised moving least squares (GMLS)~\cite{trask_high-order_2016, trask_compatible_2018,gross_meshfree_2020}, the radial basis-finite difference methods~\cite{FLYER201621}, and the local anisotropic basis function method (LABFM)~\cite{king_high_2020, king_high-order_2022, king_mesh-free_2024} are examples of mesh-free numerical methods that show high-order convergence by solving linear systems to calculate kernels or inter-particle weights. For a review of high-order mesh-free methods, we refer the reader to the introduction of ~\cite{king_high_2020}. The above methods are typically employed in an Eulerian framework, and to the best of the authors' knowledge, no formally high-order mesh-free methods have been developed and implemented at scale for a large number of degrees of freedom in a Lagrangian framework, due to the prohibitive computational costs. This creates a gap between Lagrangian and high-order mesh-free methods, which has motivated the development of computationally cheaper high-order mesh-free methods~\cite{Lind_review}.

Machine Learning (ML) has received considerable attention recently across many different fields~\cite{SHEHAB2022105458, duriez_machine_2017, KHAN20201444, Li_2022, ray_chatgpt_2023}. In particular, neural network architectures---a central component of ML---leverage the universal approximation (UA) theorem, which allows feed-forward architecture to approximate any continuous function between two Euclidean spaces~\cite{kratsios_universal_2021}. Fluid dynamics simulations, especially the ones performed in mesh-based frameworks, have been extensively enhanced using ML; for a comprehensive review, see~\cite{sharma_review_2023}. In contrast, mesh-free methods have received comparatively less attention from the ML community. Nonetheless, recent advancements have demonstrated the potential of ML in this domain, leading to notable progress. In~\cite{zago}, multi-layer perceptrons (MLPs) were used to estimate the hydrodynamic forces between particles in SPH. Results showed the MLP predictions were faithful to standard SPH results in dam break, and also presented good generalisation for different geometries than the ones present in the training dataset. In~\cite{ISHP},  graph neural networks (GNN) were implemented to surrogate the pressure Poisson's equation in incompressible SPH (ISPH). This approach produced qualitatively similar results for the tested cases, exhibited convergence behavior comparable to ISPH, and demonstrated satisfactory generalization to more complex systems. The methods discussed above leveraged the system's state to train data-driven models, which, despite demonstrating satisfactory generalization, are inherently limited to similar physical regimes. Beyond these applications, ML has also been explored as a tool to mitigate the computational costs of solving linear systems in discretized PDEs, offering alternative pathways for improving numerical efficiency.

A neural network-based algorithm to solve PDEs with high precision was developed in~\cite{jiang_neural_2023}, which solved large and sparse diagonal matrices that often arise from mesh-based numerical frameworks, and employed a residual network architecture together with a correction routine, obtaining errors smaller than $10^{-7}$ between the solution vector computed with LU factorisation and the neural network. However, it was observed that the quality of the solution deteriorated as the diagonal dominance of the linear system decreased. In~\cite{gu2023deepneuralnetworkssolving} neural networks were trained to solve large linear systems with a primary emphasis on reducing memory requirements. In this framework, the solution vector of matrix-free linear systems is approximated with a neural network representation, containing fewer degrees of freedom than the original linear system. The method proposed allows large linear systems that have prohibitive storage requirements with traditional linear solvers to be solved. However, the solution obtained with the neural network is not as accurate as traditional linear solvers, consequently the authors do not recommend the framework for small linear systems that can be solved with traditional methods. Hamrani \textit{et al.}~\cite{hamrani_machine_2023} applied ML to surrogate the computationally heavy construction of the shape function in mesh-free methods. The authors employed multiple data-driven techniques, such as gated recurrent neural networks, ensemble learning and support vector machine to predict the shape functions in the element-free Galerkin method (EFGM), radial point interpolation method (RPIM) and moving Krigin interpolation method (MKIM). The models were evaluated in one-, two- and three-dimensions, where the Gaussian process (GP) had the most accurate results, with relative errors between 0.05\% and 1.57\%, marginally outperforming MLPs. A computational cost analysis demonstrated a 20- and 1000-fold speed-up when using deep neural networks and Gaussian processes, respectively, to predict weights in three-dimensional domains compared to the original mesh-free approaches. Furthermore, a convergence analysis was conducted to compare the mesh-free algorithms with the ML-predicted weights. The neural network, Gaussian process, and the original algorithms on which they were trained exhibited similar convergence rates, with the latter producing marginally more accurate results. However, the convergence analysis was performed on coarse resolutions (less than $10^{4}$ nodes in the domain), which does not accurately reflect the resolutions encountered in standard numerical simulations. Furthermore, the work does not investigate the impact of the surrogate in differential operators, which is crucial for implementing in PDEs. Mesh-free methods may exhibit divergent behavior at high resolutions due to inconsistencies in their numerical formulation. Similarly, shape function surrogates can introduce inaccuracies, as the data-driven models often fail to perfectly predict the shape functions. Therefore, to accurately verify the stability of the ML-based shape functions, it is essential to test the surrogate across the full spectrum of relative resolutions encountered in real simulations. An ongoing issue in applied ML is the utilisation of weak baselines and reporting biased results~\cite{mcgreivy2024weakbaselinesreportingbiases,leech2024}. These biases can lead to overoptimistic assessments of model performance, which underscores the importance of rigorous and representative testing conditions to ensure the reliability and accuracy of the trained models. 

To bridge the gap between Lagrangian mesh-free simulations and Eulerian high-order mesh-free numerical methods, we propose two strategies that leverage ML to surrogate specific segments of the latter. In principle, by achieving more accurate weight computations than uncorrected SPH kernels and lower computational costs compared to high-order mesh-free methods, these surrogates can provide a accuracy computational-cost trade-off. We have tested both approaches on LABFM, given its arbitrarily high-order convergence on semi-unstructured nodes and its suitability to high-fidelity simulation of non-linear PDEs~\cite{king_mesh-free_2024}.

This paper investigates the use of MLPs to surrogate two distinct components of high-order mesh-free methods. In the first approach, a neural network is trained to directly predict the weights of a computational stencil from the positions of the support nodes relative to the central node, effectively surrogating the high-order kernel. In the second approach, a neural network is trained to solve the dense, ill-conditioned, low-rank linear systems characteristic of high-order mesh-free discretisations. Both approaches exploit the fact that the weights are exclusively a function of nodal coordinates, making the models agnostic to the underlying physics and applicable across different physical systems and mesh-free frameworks. While the first approach reveals fundamental limitations of MLPs in achieving the accuracy required for reliable convergence, the second demonstrates that a well-chosen surrogate can offer a genuine computational advantage over direct linear system solves within a specific accuracy regime.

The remainder of the paper is set out as follows. Section~\ref{sec:LABFM} and Section~\ref{sec:ML} provide a technical background on LABFM and ML, respectively. Section~\ref{sec:Method} provides a description the two approaches developed to accelerate high-order mesh-free methods. Section~\ref{sec:Results_both} presents the results obtained in the two approaches. Section~\ref{sec:caveats} provides caveats on the application of neural networks for surrogates of high-order mesh-free methods. Lastly, Section~\ref{sec:Conclusion} concludes the findings of the investigation.

\section{The Local Anisotropic Basis Function Methods (LABFM)}
In this work, we use the LABFM as an exemplar high-order mesh-free method, noting that it shares characteristics (structure of computational stencil, form of discrete operator, and local linear system) with other high-order mesh-free methods (e.g. GFDM, GMLS, radial basis function finite difference~\cite{SHU2003941, CECIL2004327}). For a comprehensive description and derivation of LABFM, we refer the reader to ~\cite{king_high_2020}.
\label{sec:LABFM}
\subsection{General Discrete Operator}
    \begin{figure}
      \centering
      \includegraphics[width=0.5\textwidth]{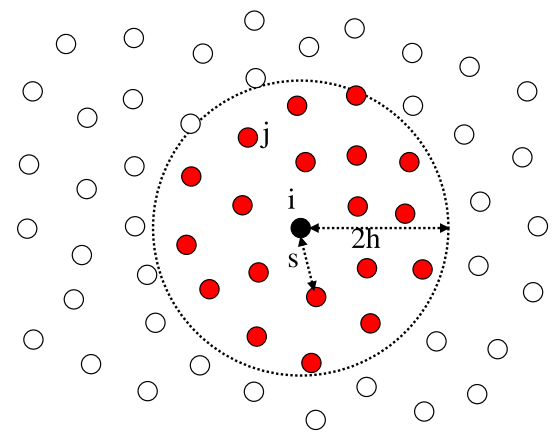}
      \caption{Illustration of a mesh-free computational stencil~\cite{king_high-order_2022}.}
      \label{fig:stecil}
    \end{figure}
In LABFM, a general discrete operator is defined to approximate differential operators as shown below:
\begin{equation}
\label{eq:discrete_op}
    L_{i}^{d}(\cdot)=\sum_{j\in\mathcal{N}}(\cdot)_{ji}w_{ji}^{d}
\end{equation}
where $L_{i}^{d}(\cdot)$ is the discrete operator; $d$ identifies the differential operator being approximated; $w_{ji}^{d}$ are a set of weights; and $\mathcal{N}$ is the set of nodes within a computational stencil; the subscripts \(i\) and \(j\) denote the central and support nodes, respectively, with scalar differences expressed as \((\cdot)_{ij} = (\cdot)_{i} - (\cdot)_{j}\). The smoothing length, \(h_{i}\), determines the size of the computational stencil, influencing the range of interactions among nodes (in~\cite{king_high_2020}, it was defined critical values for the smoothing length based on the order of approximation of the discrete operator). Figure~\ref{fig:stecil} illustrates the node distribution and stencil. The weights $w_{ji}^{d}$ are given by a weighted sum of anisotropic basis functions (ABFs),
\begin{equation}
w_{ji}^{d}=\boldsymbol{W_{ji}}\cdot\boldsymbol{\Psi_{i}^{d}}=W_{ji}^{1}\Psi_{i,1}^{d}+W_{ji}^{2}\Psi_{i,2}^{d}+W_{ji}^{3}\Psi_{i,3}^{d}+...
\end{equation}
in which the vector $\boldsymbol{W_{ji}}$ are the ABFs, and $\bm{\Psi^{d}_{i}}$ is a coefficient vector determined by solving the linear system. The coefficient vector is obtained by solving the following linear system
\begin{equation}
\label{eq:linear_system}
        \boldsymbol{M_{i}}\boldsymbol{\Psi_{i}^{d}}=\boldsymbol{C^{d}}
\end{equation}
The vector, $\boldsymbol{C^{d}}$ (also referred to as the stencil moments), ensures that the coefficient vector, $\boldsymbol{\Psi_{i}^{d}}$, is appropriate to approximate the desired differential operator. For first-order spatial derivatives and the Laplace operator in a two-dimensional domain the are given by:
\begin{equation}
\label{eq:labfm_moments}
        \boldsymbol{C^{d}}=
        \begin{cases} 
            \text{$[1, 0, 0, 0, 0, 0, ...]^{T}$} & \text{$if$ $d=x$} \\
            \text{$[0, 1, 0, 0, 0, 0, ...]^{T}$} & \text{$if$ $d=y$} \\
            \text{$[0, 0, 1, 0, 1, 0, ...]^{T}$} & \text{$if$ $d=\Delta$}
        \end{cases} 
\end{equation}
where $x$, $y$ and $\Delta$ refer respectively to the spatial derivatives in each respective direction and the Laplace operator. For a given support region, $\boldsymbol{M_{i}}$ is independent of the differential operator being approximated, given by:
\begin{equation}
\label{eq:lin_sys}    \boldsymbol{M_{i}}=\sum_{j\in\mathcal{N}}\boldsymbol{X_{ji}}\otimes\boldsymbol{W_{ji}}
\end{equation}
where $\boldsymbol{X_{ji}}$ is the vector containing Taylor monomials
\begin{equation}
        \boldsymbol{X_{ji}}=\left[x_{ji}, y_{ji}, \frac{x_{ji}^{2}}{2!}, x_{ji}y_{ji}, \frac{y_{ji}^{2}}{2!}, \frac{x_{ji}^3}{3!}, \frac{x_{ji}^{2}y_{ji}}{2!}, ...\right]^{T}
\end{equation}
note that for the monomial vector $\boldsymbol{X_{ji}}\neq\boldsymbol{X_{j}}-\boldsymbol{X_{i}}$. The matrix $\bm{M_{i}}$ as defined above has infinite rank. In practice, we restrict the size of $\bm{M}_{i}$ by including only the first $p$ elements of $\bm{W_{ji}}$ and $\bm{X_{ji}}$. We highlight here the relationship between the number of elements $p$, and the order $k$ of the $p$-th element of $\boldsymbol{X_{ji}}$
\begin{equation}
    p = \frac{k^2+3k}{2} \quad
\end{equation}
in 2 dimensions.

\subsection{Anisotropic Basis Functions}
To construct the ABFs, bivariate Hermite polynomials are combined with a radial basis function (RBF). The $q$-th element of $\boldsymbol{X}_{ji}$ is proportional to $x_{ji}^{a}y_{ji}^{b}$. Thus, the $q$-th ABF is defined as
\begin{equation}
\label{eq:abf}
        W_{ji}^{q}=\frac{\psi\left(|\boldsymbol{r}_{ji}|/h_{i}\right)}{\sqrt{2^{a+b}}}H_{a}\left(\frac{x_{ji}}{h_{i}\sqrt{2}}\right)H_{b}\left(\frac{y_{ji}}{h_{i}\sqrt{2}}\right)
\end{equation}
where $\boldsymbol{r}_{ji}$ is the Euclidean distance between the nodes $j$ and $i$; $H_{a}$ is the $a$-th order univariate Hermite polynomial (of the physics kind), generated from
    \begin{equation}
        H_{a}(z)=(-1)^{a}e^{z^{2}}\frac{d^{a}}{dz^{a}}e^{-z^{2}}
    \end{equation}
    and $\psi$ is an RBF. In this work, the Wendland C2 kernel (shown in Appendix~\ref{appendix:wendland} is used, following~\cite{king_high-order_2022}).

\section{Machine Learning}
\label{sec:ML}
\subsection{Multi-Layer Perceptrons}
\label{sec:feed_forward}
The foundational structures of MLPs consist of an input layer, hidden layers, and an output layer. The hidden layers consist of perceptrons (or neurons) interconnected by weights across successive layers. Each neuron receives input from the preceding layer and, in the most basic architectures, performs two primary functions: calculating a weighted sum of its inputs and applying an activation function to the resulting sum. The operations in each layer can be expressed as follows:
    \begin{subequations}
    \begin{equation}
    \label{eq:neuron1}
        \mathbf{c}^{l} = \boldsymbol{\theta}^{l} \mathbf{a}^{l-1} + \mathbf{b}^{l}
    \end{equation}
    \begin{equation}
    \label{eq:neuron2}
        \mathbf{a}^{l} = \sigma\left(\mathbf{c}^{l}\right)
    \end{equation}
    \end{subequations}
where superscript $l$ refers to the layer index; $\mathbf{c}$ represents the pre-activation vector, which is the weighted sum of the activations of the previous layer; $\mathbf{a}$ refers to the vector of activations; $\mathbf{b}$ represents the bias vector, and $\boldsymbol{\theta}$ is the weight matrix that connects the neurons in successive layers; $\sigma$ is the activation function, which introduces non-linearity into the network, enabling it to exhibit the universal approximation (UA) property. Examples of activation functions are the sigmoid and rectified linear unit (ReLU).  For a detailed review of MLPs, the reader is referred to~\cite{Goodfellow-et-al-2016}.

\section{Method}
In the current section, we will demonstrate two approaches that surrogate segments of high-order mesh-free methods with ML. These approaches were implemented in LABFM as an exemplar, due to its high-order convergence and applicability to high-fidelity simulation of non-linear PDEs. The objective is to approximate differential operators with high-order accuracy and increased computational efficiency to bridge the existing gap between Lagrangian mesh-free methods and high-order interpolants.
\label{sec:Method}

\subsection{High-Order Kernel Surrogate with Machine Learning}
\label{sec:Kernel surr w ML}
In this investigation, we leveraged the UA property of feed-forward neural networks to map the distance between the support nodes and the central node to the LABFM weights, \( w_{ji}^{d} \). The network inputs were the positions of each support node $j\in\mathcal{N}$ relative to the central node $i$, given by the sets \( \{ x_{ji} \}_{j \in \mathcal{N}} \) and \( \{ y_{ji} \}_{j \in \mathcal{N}} \), ordered by radial distance. The central node and its corresponding weight were excluded from the model. This exclusion is justified by the fact that, according to Equation~\eqref{eq:discrete_op}, the weight of the central node is irrelevant, since \( (\cdot)_{i} - (\cdot)_{i} = 0 \), it has no impact on the discrete operator. Consequently, the input vector consists only of the coordinate differences for the support nodes, resulting in a size of \( \mathbb{R}^{d|\mathcal{N}| - d} \), and the output consists of the LABFM weights for the support nodes, with a size of \( \mathbb{R}^{|\mathcal{N}| -1} \), where $d$ is the number of dimensions.

The inflexibility of feed-forward networks with varying input and output size requires a fixed number of support nodes in the support region. In LABFM, the number of support nodes must be $|\mathcal{N}| \geq |\mathcal{N}_{crit}|$, where $|\mathcal{N}_{crit}|$ was defined in King \textit{et al.}~\cite{king_high_2020} for different orders of approximation. In this work, we use LABFM at second-order, which is expected to be easier to surrogate compared to higher-order approximations. Higher-order LABFM approximations introduce high-frequency variations in the weights, particularly near the stencil edges, which conflicts with the spectral bias of artificial neural networks toward learning smooth, low-frequency functions. Each computational stencil contains 30 support nodes, matching the approximate neighbour count of the SPH Wendland C2 kernel (Appendix~\ref{app:sph}).

Proper normalisation of inputs and outputs improves weight convergence during network trainings~\cite{dnn_normalisation}. In the current investigation, the training dataset comprises points from domains with varying resolution, leading to significant variations in distances and weight magnitudes. These weights are inversely proportional to the smoothing length of the computational stencil, $w \propto 1/h$ and $w \propto 1/h^2$ for first and second derivatives, respectively. To mitigate resolution-dependent variability, weights and distances are normalized by the smoothing length of each computational stencil. This normalization rescales the weights to a uniform magnitude, independent of the stencil size and the underlying resolution, ensuring consistent learning and prediction across different node resolutions, as detailed below:
    \begin{subequations}
    \label{eq:hok_norm}
    \begin{equation}
        \hat{x}_{ij} = \frac{x_{ij}}{h_i}, \quad \hat{y}_{ij} = \frac{y_{ij}}{h_i}
    \end{equation}
    \begin{equation}
    \hat{w}_{ij} = 
    \begin{cases} 
        \frac{w_{ij}}{h_{i}} & \text{if } d = x, y \\
        \frac{w_{ij}}{h_{i}^2} & \text{if } d = L
    \end{cases} 
    \end{equation}
    \end{subequations}

\subsection{Solving Linear Systems with Machine Learning}
\label{sec:Linear sys w ML}
The second approach developed in this paper is a more direct way to accelerate high-order mesh-free methods with ML. The computational overhead in high-order interpolants is largely caused by the linear systems that must be solved to calculate weights. Thus, in this approach, we implement feed-forward networks to surrogate the solution of the dense, ill-conditioned, low-rank linear systems with a constant right-hand side vector (given by~\eqref{eq:linear_system} in LABFM), which contains fewer intermediate steps compared to the approach shown in Section~\ref{sec:Kernel surr w ML}.

The models built in this approach have as input the flattened $\boldsymbol{M_{i}}$ matrix and as output the weights, $\boldsymbol{\Psi_{i}^{d}}$. The network operations can be described as $f : \mathbb{R}^{k^{2}} \rightarrow \mathbb{R}^{k}$, where $k$ is the rank of the linear system. Given the moments are constant for each differential operator being approximated, they do not provide additional information and can be omitted from the network. 

During training, the moments were normalised with respect to the smoothing length, $h_i$. Similarly to the approach presented in Section~\ref{sec:Kernel surr w ML}, the normalisation depends on which operator is being approximated. 
    \begin{equation}
        \hat{\Psi}_{i} = 
        \begin{cases} 
            \text{$\frac{\Psi_{i}}{h_{i}}$} & \text{$if$ $d=x,y$} \\
            \text{$\frac{\Psi_{i}}{h_{i}^{2}}$} & \text{$if$ $d=L$}
        \end{cases} 
    \end{equation}
This normalization rescales the weights to a magnitude independent of node distance, ensuring learning and prediction across different node resolutions. The entries of the linear system for a second-order approximation exhibit uniform magnitudes and are relatively small. Consequently, normalization of the input data during training was deemed unnecessary.

\subsection{Training \& Testing Details}
Training data consists of local particle neighbourhoods $\mathcal{N}_i$ generated from synthetic disordered point clouds. Starting from a regular Cartesian grid with mean spacing $s_u$, each node is independently perturbed by a displacement drawn from $\mathcal{U}(-\frac{\epsilon s_u}{2}, \frac{\epsilon s_u}{2})$ along each coordinate, where $\epsilon$ is the non-dimensional disorder intensity. By construction, the perturbation magnitude scales with the grid spacing, allowing operator sensitivity to be studied systematically across discretisation scales. Unless stated otherwise, all models are trained at $\epsilon = 1.0$, a disorder level considerably higher than that encountered in typical mesh-free simulations, ensuring robustness under severe geometric irregularity.

The MLPs were trained using the PyTorch framework~\cite{pytorch} with a mean squared error loss function. The training dataset comprised approximately 1.6 million computational stencils, split 80\% for training and 20\% for validation. Throughout Section~\ref{sec:Results_both}, superscripts in the kernel name denote the target operator (e.g.\ the Laplacian) and subscripts denote the order of accuracy used to generate the LABFM weights (e.g.\ LABFM$_{p=2}$ indicates second-order weights).

The networks underwent offline training and their results were evaluated in three distinct phases. The first is a qualitative analysis of the weights. The second phase involved using the networks to compute the stencil moments with the predicted weights, a critical measure of the quality of the predicted weights. Lastly, a convergence study was conducted using the ML-computed weights with the designated test function:
    \begin{equation}
    \label{eq:test_func}
        \phi(x^{*},y^{*})=1.0 + (x^{*}y^{*})^4 + \sum_{i=1}^{6}(x^{*i}+y^{*i})
    \end{equation}
where $x^{*} = x - 0.1453$ and $y^{*}=y - 0.16401$. This equation matches the one used in \cite{king_high_2020}. The offset is pseudo-random, and included to ensure asymmetry in the function, to prevent the masking of errors which could cancel for a symmetric function. 

\section{Results \& Discussion}
\label{sec:Results_both}
\subsection{Predicted Weighted Anisotropic Basis Functions}
In this section, we present results for surrogating the high-order kernel with an MLP. The analysis is divided into two parts: (i) a qualitative analysis (Section~\ref{sec:r1_qual}) and (ii) a quantitative analysis (Section~\ref{sec:r1_quant}).

\label{sec:Results1}
\subsubsection{Qualitative analysis}
\label{sec:r1_qual}
\begin{figure}[ht]
\begin{center}
  \begin{subfigure}[t]{0.24\linewidth}
    \begin{center}
      \includegraphics[width=\linewidth]{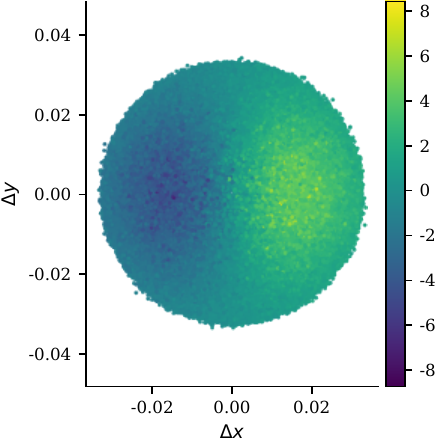}
    \end{center}
  \end{subfigure}\hfill
  \begin{subfigure}[t]{0.24\linewidth}
    \begin{center}
      \includegraphics[width=\linewidth]{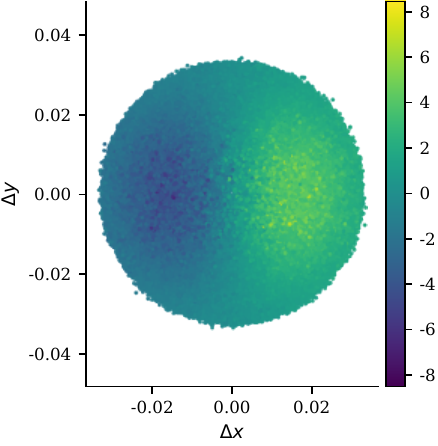}
    \end{center}
  \end{subfigure}\hfill
  \begin{subfigure}[t]{0.24\linewidth}
    \begin{center}
      \includegraphics[width=\linewidth]{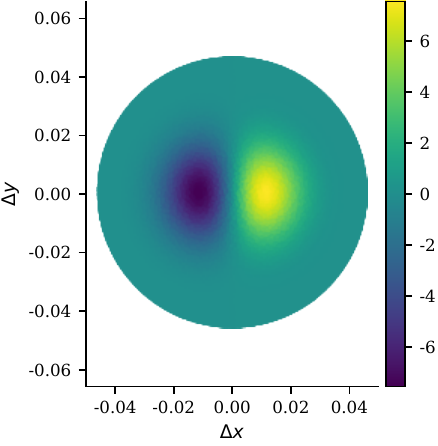}
    \end{center}
  \end{subfigure}\hfill
  \begin{subfigure}[t]{0.24\linewidth}
    \begin{center}
      \includegraphics[width=\linewidth]{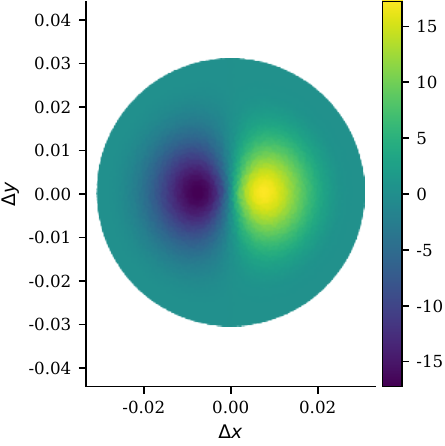}
    \end{center}
  \end{subfigure}

  \vspace{1em}

  \begin{subfigure}[t]{0.24\linewidth}
    \begin{center}
      \includegraphics[width=\linewidth]{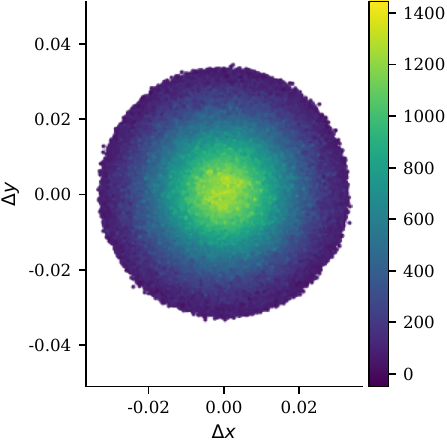}
    \end{center}
    \caption{LABFM}
  \end{subfigure}\hfill
  \begin{subfigure}[t]{0.24\linewidth}
    \begin{center}
      \includegraphics[width=\linewidth]{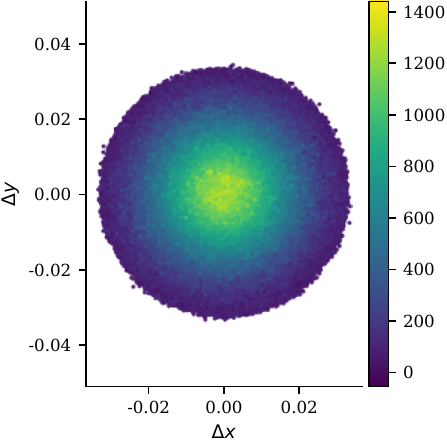}
    \end{center}
    \caption{MLP}
  \end{subfigure}\hfill
  \begin{subfigure}[t]{0.24\linewidth}
    \begin{center}
      \includegraphics[width=\linewidth]{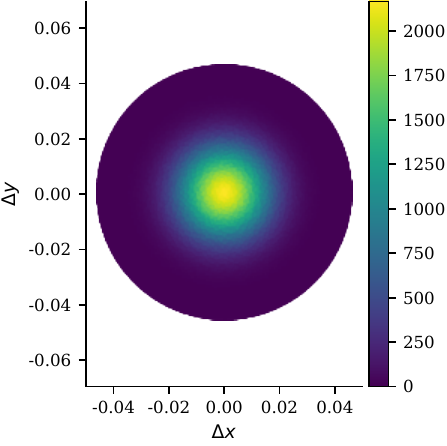}
    \end{center}
    \caption{Quintic Spline}
  \end{subfigure}\hfill
  \begin{subfigure}[t]{0.24\linewidth}
    \begin{center}
      \includegraphics[width=\linewidth]{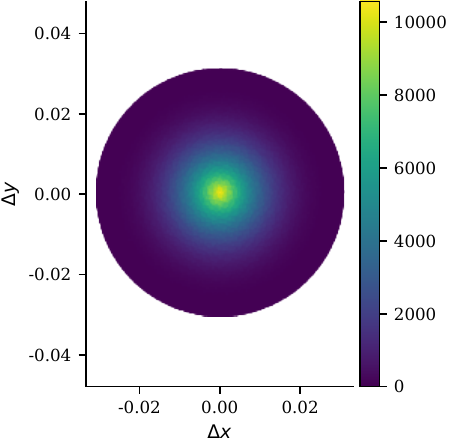}
    \end{center}
    \caption{Wendland C2}
  \end{subfigure}
\end{center}
\caption{Overlaid plot of multiple computational stencils with different discretisation methods for $x$-derivative (\textit{top row}) and Laplacian (\textit{bottom row}), colours indicate weight value. The MLPs were trained with particle disturbance of $\epsilon=1.0$ and inferred with the same noise.}
\label{fig:r1_k}
\end{figure}

Figure~\ref{fig:r1_k} shows results for the LABFM, MLP surrogate, quintic spline, and Wendland C2 kernels evaluated on multiple computational stencils absent from the training dataset. The stencils are overlaid, with colour representing weight magnitude. The top row shows the weights predicted for the $x$-derivative operator and the bottom row for the Laplacian. Visually, LABFM and the MLP surrogate are indistinguishable, exhibiting similar weight magnitudes and distributions across the stencil ensemble. In contrast, the SPH kernels produce considerably smoother weight fields. This is because SPH weights depend only on the relative position between each neighbour and the central node, making each weight assignment independent of the remaining nodes in the stencil. LABFM, on the other hand, computes weights by solving a linear system constructed from the positions of all nodes within the stencil (Equation~\eqref{eq:lin_sys}), which is smooth for each individual stencil but appears noisy when superimposed. Despite this apparent noise, for the $x$-derivative, both LABFM and the MLP surrogate exhibit approximate anti-symmetry in the weight distribution about the origin, in the direction of the derivative, and rotational symmetry for the Laplacian. The stencil boundaries also appear noisier than those of the SPH kernels, which is attributed to the use of a fixed neighbour count in LABFM and the MLP surrogate, as opposed to the fixed cutoff radius employed by SPH kernels. Although LABFM does not require a fixed neighbour count, it was set to match the MLP surrogate for a fair comparison.

\subsubsection{Quantitative Analysis}
\label{sec:r1_quant}
Quantitatively, the results are evaluated through stencil moments and convergence analysis. For all stencils in the validation dataset, the predicted weights were used to compute the moments of the $x$-derivative and Laplacian operators. The absolute error between predicted and target moments was then computed and averaged across all stencils, with the standard deviation also reported. These residuals indicate the precision to which each term in the Taylor expansion approximates the true differential operator, up to the truncated order. The absolute error between the target and the predicted stencil moments yields the linear system residual in Equation~\eqref{eq:linear_system}. The results are summarised in Table~\ref{table:labfm_moments_1}.

\begin{table}[t]
\caption{Moment residuals for MLP and SPH discretisations. Mean absolute error (MAE) and standard deviation are reported for each monomial moment, averaged over independently perturbed neighbourhood realisations with $\epsilon=1.0$. Residuals quantify the extent to which the discretisation satisfy the imposed Taylor consistency constraints. Superscripts denote the target operator and subscripts the polynomial order of approximation.}
\label{table:labfm_moments_1}
\begin{center}
\setlength{\tabcolsep}{8pt} 
\renewcommand{\arraystretch}{1.1} 
\begin{tabular}{ccccccc}
\multicolumn{1}{c}{\bf Kernel} &
\multicolumn{1}{c}{\bf Metric} &
\multicolumn{1}{c}{\bf $x$} &
\multicolumn{1}{c}{\bf $y$} &
\multicolumn{1}{c}{\bf $x^2/2$} &
\multicolumn{1}{c}{\bf $xy$} &
\multicolumn{1}{c}{\bf $y^2/2$}
\\ \hline \\

\multirow{2}{*}{$\text{MLP}_{p=2}^x$}& MAE  & $2.87 \times 10^{-3}$ & $5.18 \times 10^{-2}$ & $1.44 \times 10^{-3}$ & $3.18 \times 10^{-2}$ & $4.93 \times 10^{-3}$ \\
                      & St.d. & $3.75 \times 10^{-3}$ & $6.52 \times 10^{-2}$ & $1.88 \times 10^{-3}$ & $4.00 \times 10^{-2}$ & $6.26 \times 10^{-3}$ \\
\hline

\multirow{2}{*}{$\text{MLP}_{p=2}^\Delta$} & MAE   & $3.84 \times 10^{-2}$ & $3.79 \times 10^{-2}$ & $9.11 \times 10^{-3}$ & $1.74 \times 10^{-2}$ & $9.67 \times 10^{-3}$ \\
                           & St.d. & $5.03 \times 10^{-2}$ & $4.97 \times 10^{-2}$ & $1.19 \times 10^{-2}$ & $2.27 \times 10^{-2}$ & $1.18 \times 10^{-2}$ \\
\hline

\multirow{2}{*}{$\text{Wendland C2}^x$} & MAE   & $1.01 \times 10^{-1}$ & $7.08 \times 10^{-2}$ & $4.80 \times 10^{-2}$ & $4.93 \times 10^{-2}$ & $2.55 \times 10^{-2}$ \\
                           & St.d. & $1.21 \times 10^{-1}$ & $8.84 \times 10^{-2}$ & $5.90 \times 10^{-2}$ & $6.16 \times 10^{-2}$ & $3.18 \times 10^{-2}$ \\
                           \hline
\multirow{2}{*}{$\text{Wendland C2}^\Delta$} & MAE   & $3.31 \times 10^{-1}$ & $3.33 \times 10^{-1}$ & $1.03 \times 10^{-1}$ & $1.50 \times 10^{-1}$ & $1.05 \times 10^{-1}$ \\
                           & St.d. & $4.13 \times 10^{-1}$ & $4.13 \times 10^{-1}$ & $1.22 \times 10^{-1}$ & $1.87 \times 10^{-1}$ & $1.24 \times 10^{-1}$ \\
\hline
\multirow{2}{*}{$\text{Quintic Spline}^x$} & MAE   & $6.59 \times 10^{-2}$ & $4.57 \times 10^{-2}$ & $3.77 \times 10^{-2}$ & $3.98 \times 10^{-2}$ & $2.12 \times 10^{-2}$ \\
                           & St.d. & $8.04 \times 10^{-2}$ & $5.75 \times 10^{-2}$ & $4.69 \times 10^{-2}$ & $4.95 \times 10^{-2}$ & $2.64 \times 10^{-2}$ \\
                           \hline

\multirow{2}{*}{$\text{Quintic Spline}^\Delta$} & MAE   & $1.52 \times 10^{-1}$ & $1.49 \times 10^{-1}$ & $6.29 \times 10^{-2}$ & $9.47 \times 10^{-2}$ & $6.42 \times 10^{-2}$ \\
                           & St.d. & $1.90 \times 10^{-1}$ & $1.86 \times 10^{-1}$ & $7.70 \times 10^{-2}$ & $1.18 \times 10^{-1}$ & $7.89 \times 10^{-2}$\\

\end{tabular}
\end{center}
\end{table}

Regarding the gradient operator, MLP residuals remain consistently in the range $error\in\left[\mathcal{O}(10^{-2}),\mathcal{O}(10^{-3})\right]$ across all first- and second-order monomials, compared to SPH residuals of $error\in\left[\mathcal{O}(10^{-1}),\mathcal{O}(10^{-2})\right]$. The MLP Laplacian yields slightly higher residuals than the derivative operator with residuals mostly of order $10^{-2}$, whereas for the SPH kernels the gap is more pronounced, particularly in the first-order moments. This degradation in the Laplacian is consistent with the heightened sensitivity of second-order derivatives to geometric perturbations, a well-documented behaviour in high-order mesh-free and spectral discretisations~\cite{Lin2025ISPH}. Taken together, these results suggest that polynomial consistency constraints can be learned within a supervised framework, and generalise to unseen stencil configurations. However, the accuracy obtained is fairly limited, given the learned weights marginally outperform uncorrected SPH.

\begin{figure}[t]
  \begin{center}
    \begin{subfigure}{0.49\linewidth}
      \centering
      \includegraphics[width=\linewidth]{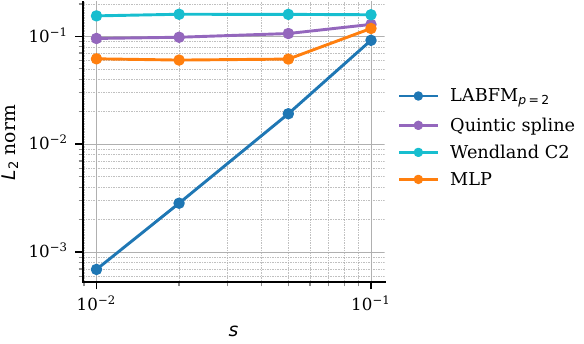}
    \end{subfigure}
    \begin{subfigure}{0.49\linewidth}
      \centering
      \includegraphics[width=\linewidth]{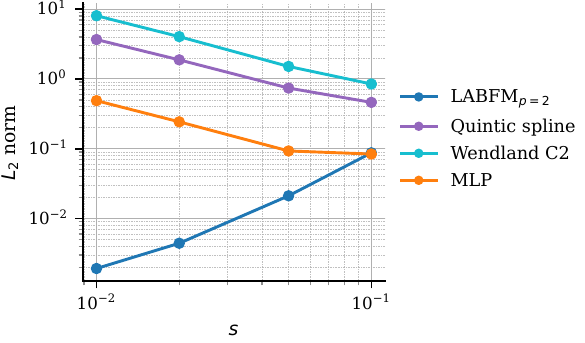}
    \end{subfigure}
  \end{center}
  \caption{Convergence of the $x$-derivative (\textit{left}) and Laplacian (\textit{right}) on a smooth test function with $\epsilon = 1.0$. Relative $L_2$ error versus particle spacing $s$ for the MLP as a discretisation surrogate, a second-order LABFM kernel, and two uncorrected SPH kernels (quintic spline and Wendland C2).}
  \label{fig:convergence}
\end{figure}

A convergence analysis was subsequently performed using the trained models to approximate the derivative and Laplacian operators via a test function (Equation~\ref{eq:test_func}), with results shown in Figure~\ref{fig:convergence}. For the derivative operator, the MLP surrogate exhibits convergence only for  $s \in [0.1,0.5]$, marginally outperforming the SPH kernels. The limiting error in the convergence is governed by the sum of the first-order moment residual. Outside this range, the MLP surrogate fails to converge, likely because the moment residuals reported in Table~\ref{table:labfm_moments_1} are insufficient to sustain consistency at finer resolutions. LABFM, by contrast, achieves second-order convergence across the entire resolution range tested, as consistency is enforced up to the truncation error of the linear system solve~\cite{shankar2014radialbasisfunctionrbffinite, king_high_2020}. For the Laplacian, both the SPH and MLP operators exhibit divergent behaviour across all resolutions tested, which follows directly from the failure to satisfy the moment conditions to a sufficient degree, as evidenced by the higher residuals in Table~\ref{table:labfm_moments_1}. These convergence results expose a fundamental limitation of directly surrogating the high-order kernel: although the MLP learns a qualitatively accurate approximation, polynomial consistency is highly sensitive to weight errors, and the residual moment errors are large enough to preclude convergence.

\subsection{Analysis of Machine Learning Solution of Linear Systems}
\label{sec:Results2}
In this section, we present results for surrogating the ill-conditioned, low-rank, dense linear systems with constant right-hand side vectors present in high-order mesh-free methods with a MLP.

The predicted kernel qualitatively matches the LABFM kernel solved by LU factorisation (Figure~\ref{fig:r1_k}), for both the derivative and Laplacian operators, exhibiting similar weight magnitudes, anti-symmetric distributions for the derivative, and rotational symmetry for the Laplacian. The same qualitative features discussed in Section~\ref{sec:r1_qual} are observed here, and the same interpretation applies.

\begin{figure}[t]
  \begin{center}
    \begin{subfigure}{0.4\linewidth}
      \centering
      \includegraphics[width=\linewidth]{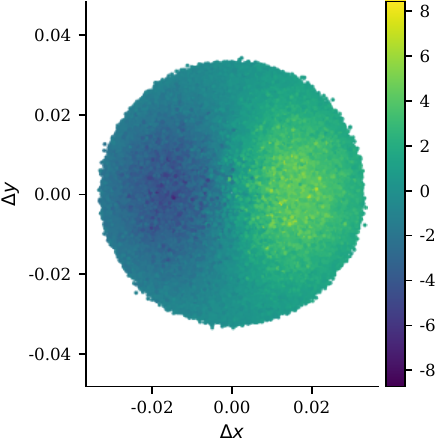}
    \end{subfigure}
    \begin{subfigure}{0.4\linewidth}
      \centering
      \includegraphics[width=\linewidth]{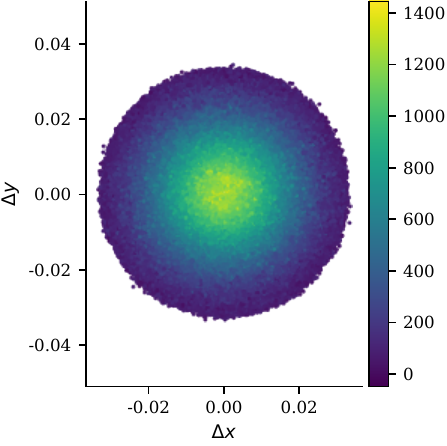}
    \end{subfigure}
  \end{center}
\caption{Overlaid plot of multiple computational stencils with MLP surrogating the linear system for $x$-derivative (\textit{left}) and Laplacian (\textit{right}), colours indicate weight value. The MLPs were trained with particle disturbance of $\epsilon=1.0$ and inferred with the same noise.}
\label{fig:r2_k}
\end{figure}

Following Section~\ref{sec:r1_quant}, we evaluate the consistency and convergence of the linear system surrogate, with moment residuals reported in Table~\ref{table:labfm_moments_2}. For the derivative operator, first-order moment residuals are of order $10^{-4}$ across all models, while second-order moments achieve slightly higher accuracy at $10^{-5}$, for both mean absolute error and standard deviation. The Laplacian exhibits marginally larger residuals than the $x$-derivative, consistent with the trend observed in Table~\ref{table:labfm_moments_1}. Overall, surrogating the linear system yields a reduction of two to three orders of magnitude in moment residuals compared to directly surrogating the kernel, representing a substantial improvement in polynomial consistency.

\begin{table}[t]
\caption{Moment residuals for MLP linear system surrogate. Mean absolute error (MAE) and standard deviation are reported for each monomial moment, averaged over independently perturbed neighbourhood realisations with $\epsilon=1.0$. Residuals quantify the extent to which the discretisation satisfy the imposed Taylor consistency constraints. Superscripts denote the target operator and subscripts the polynomial order of approximation.}

\label{table:labfm_moments_2}
\begin{center}
\setlength{\tabcolsep}{8pt} 
\renewcommand{\arraystretch}{1.1} 
\begin{tabular}{ccccccc}
\multicolumn{1}{c}{\bf Operator} &
\multicolumn{1}{c}{\bf Metric} &
\multicolumn{1}{c}{\bf $x$} &
\multicolumn{1}{c}{\bf $y$} &
\multicolumn{1}{c}{\bf $x^2/2$} &
\multicolumn{1}{c}{\bf $xy$} &
\multicolumn{1}{c}{\bf $y^2/2$}
\\ \hline \\

\multirow{2}{*}{$\text{MLP}_{p=2}^x$}& MAE  & $1.66 \times 10^{-4}$ & $1.28 \times 10^{-4}$ & $5.40 \times 10^{-5}$ & $2.54 \times 10^{-5}$ & $5.45 \times 10^{-5}$ \\
                      & St.d. & $2.11 \times 10^{-4}$ & $2.05 \times 10^{-4}$ & $8.62 \times 10^{-5}$ & $4.22 \times 10^{-5}$ & $8.56 \times 10^{-5}$ \\
\hline

\multirow{2}{*}{$\text{MLP}_{p=2}^\Delta$} & MAE   & $3.18 \times 10^{-4}$ & $3.13 \times 10^{-4}$ & $1.77 \times 10^{-4}$ & $7.03 \times 10^{-5}$ & $1.84 \times 10^{-4}$ \\
                           & St.d. & $5.03 \times 10^{-4}$ & $5.23 \times 10^{-4}$ & $2.75 \times 10^{-4}$ & $1.26 \times 10^{-4}$ & $2.79 \times 10^{-4}$ \\
\hline

\end{tabular}
\end{center}
\end{table}

\begin{figure}[t]
  \begin{center}
    \begin{subfigure}{0.49\linewidth}
      \centering
      \includegraphics[width=\linewidth]{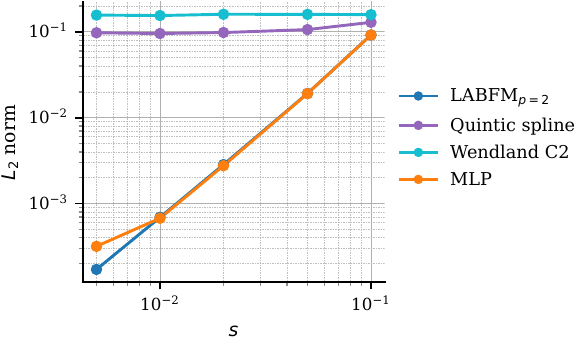}
    \end{subfigure}
    \begin{subfigure}{0.49\linewidth}
      \centering
      \includegraphics[width=\linewidth]{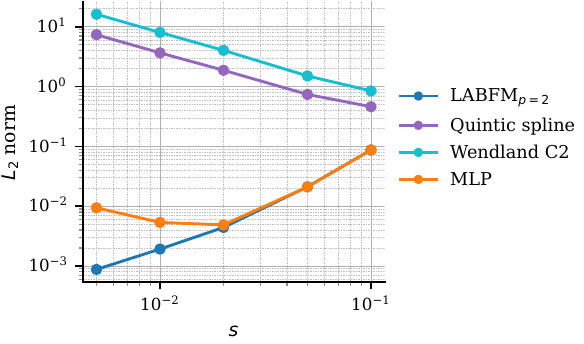}
    \end{subfigure}
  \end{center}
  \caption{Convergence of the $x$-derivative (\textit{left}) and Laplacian (\textit{right}) on a smooth test function with $\epsilon = 1.0$. Relative $L_2$ error versus particle spacing $s$ for the MLP as a linear system surrogate, a second-order LABFM kernel with LU factorisation, and two uncorrected SPH kernels (quintic spline and Wendland C2).}
  \label{fig:convergence2}
\end{figure}

A convergence analysis was subsequently performed comparing the linear system surrogate against standard LABFM with LU factorisation and the SPH kernels, with results shown in Figure~\ref{fig:convergence2}. For the $x$-derivative, the MLP surrogate exactly matches LABFM convergence up to $s=10^{-2}$, beyond which it plateaus, indicating that the limiting error has been reached. The Laplacian exhibits similar behaviour, with the surrogate matching LABFM up to a lower resolution and at reduced accuracy compared to the derivative operator, before diverging. Overall, this approach represents a marked improvement over the full kernel surrogate. While the limiting error is necessarily larger than LABFM — since the linear system is not solved explicitly — the results demonstrate that the surrogate is viable for applications requiring moderate accuracy.

Lastly, a computational cost analysis was conducted to assess the competitiveness of the MLP surrogate relative to LU factorisation for solving the linear system. Since the output must be rescaled to its correct magnitude before use, the total forward pass time is reported. Two models are evaluated: MLP$_l$, a larger model with more trainable parameters targeting a lower limiting error, and MLP$_s$, a smaller model prioritising computational efficiency. 

    \begin{figure}
      \centering
      \includegraphics[width=0.65\textwidth]{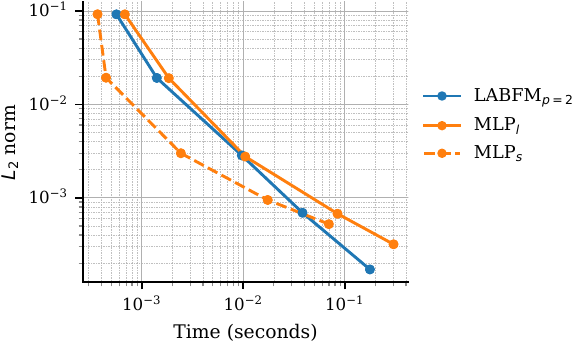}
      \caption{Evaluation of the trade-off between parameter count, wall-clock forward time, and accuracy for the MLP linear system surrogate and baseline LABFM when computing stencil weights. The vertical axis shows the $L_2$ error of the $x$-derivative of the test function in Equation~\eqref{eq:test_func} at different resolutions.}
      \label{fig:comp_cost}
    \end{figure}
    
Figure~\ref{fig:comp_cost} shows the accuracy--computational cost trade-off for the linear system MLP surrogates. MLP$_l$ is consistently more expensive than both LABFM and MLP$_s$, though it maintains convergent behaviour across all resolutions tested, reflecting the accuracy benefit of increased model capacity. MLP$_s$ outperforms LABFM in computational cost over a moderate resolution range, achieving up to a $5\times$ speedup, but eventually plateaus at its limiting error as resolution increases. Within this range, MLP$_s$ achieves comparable accuracy to LABFM at lower cost, demonstrating an efficiency advantage for applications requiring moderate accuracy with a second-order LABFM discretisation. These results, however, are specific to the second-order linear system; extending to higher-order approximations introduces additional challenges, as the conditioning of the linear system deteriorates with increasing polynomial order, which we discuss next.

\section{Caveats Regarding High-Order Approximation Accuracy}
\label{sec:caveats}
LABFM has a theoretical leading order error of $\mathcal{O}(s_{i}^{k+1-\zeta})$, where $\zeta$ represents the order of the approximated differential operator, and $k$ is the order of approximation of the differential operator. However, the conditioning of the linear system (Equation~\ref{eq:linear_system}) deteriorates for higher-order approximations, amplifying the impact of numerical errors and increasing the sensitivity of the solution vector to noise; consequently, there is an increase in the limiting error for higher-order approximations even with standard linear system solvers such as LU factorisation~\cite{king_high-order_2022}. 

\begin{figure}
      \centering
      \includegraphics[width=0.7\textwidth]{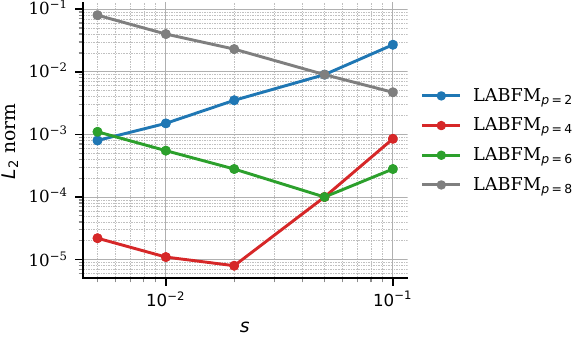}
      \caption{Convergence study of LABFM approximation of the Laplace operator of Equation~\eqref{eq:test_func} for different orders of approximation with noise introduced in coefficient vector ($\boldsymbol{\Psi_{i}^d}$).}
      \label{fig:noisy_convergence}
    \end{figure}

To assess the sensitivity of the LABFM operator to errors in the solution vector at higher orders of approximation, we define a noise bound as $\eta \|\boldsymbol{\Psi_i^d}\|_\infty$, where $\eta = 10^{-7}$. Independent noise values are sampled uniformly from $\mathcal{U}\left(-\eta \|\boldsymbol{\Psi_i^d}\|_\infty,\, \eta \|\boldsymbol{\Psi_i^d}\|_\infty\right)$ and added to each entry of the solution vector $\boldsymbol{\Psi_i^d}$. This scaling ensures the injected noise is proportional to the solution magnitude while remaining well below the error levels achieved by any model in this study. A convergence analysis is then performed with the noise-perturbed solution vector, with results shown in Figure~\ref{fig:noisy_convergence}.

As shown in Figure~\ref{fig:noisy_convergence}, even minimal noise in the solution vector is sufficient to disrupt convergence at higher orders, consistent with the conditioning deterioration reported in~\cite{king_high-order_2022}. The second-order approximation remains convergent across all simulated resolutions. The fourth-order approximation diverges at a resolution of $2\times10^{-2}$, with a limiting error of $\sim10^{-5}$, while the sixth-order diverges earlier, at $5\times10^{-2}$, with a limiting error of $\sim10^{-4}$. The eighth-order approximation exhibits divergent behaviour throughout the entire resolution range tested. The progressive degradation in both limiting error and onset resolution confirms that higher-order approximations impose increasingly stringent accuracy requirements on the predicted solution vector.

This sensitivity is a direct consequence of the higher-order moment constraints. As the approximation order increases, the solution vector must satisfy additional polynomial terms in the Taylor expansion, rendering the underlying optimisation problem increasingly stiff and the required prediction accuracy increasingly difficult to achieve with current neural network architectures.

A further fundamental obstacle is the well-documented spectral bias of neural networks~\cite{spectralbiasneuralnetworks,Zhi_Qin_John_Xu_2020}, which causes them to converge more slowly on high-frequency modes. The qualitative results in Figures~\ref{fig:r1_k}-~\ref{fig:r2_k} confirm that the dominant low-frequency components were accurately learned, producing visually indistinguishable weight fields. However, the residual error shown in Tables~\ref{table:labfm_moments_1}-~\ref{table:labfm_moments_2} reveal that the network only partially captures the high-frequency residuals, which are precisely the components that govern moment consistency at fine resolutions. This is compounded by intrinsic limitations of gradient-based training algorithms~\cite{Boche_2023}: Colbrook \textit{et al.}~\cite{nn_limit_accuracy} rigorously demonstrate that even when a stable and accurate network exists in theory, training algorithms face fundamental computational barriers beyond a certain precision threshold, with further improvements requiring disproportionately more data or being unattainable in the worst case. This provides a principled explanation for the error floors observed in both surrogate approaches.

Beyond accuracy, practical deployment introduces a competing constraint: prediction speed. Achieving lower error floors generally demands larger, more computationally expensive models, directly conflicting with the efficiency motivation for surrogating the linear system solve. This trade-off fundamentally limits the operating regime of standard multi-layer perceptrons surrogates for high-order discretisations. 

\section{Summary \& Conclusion}
\label{sec:Conclusion}
Mesh-free numerical methods have gained considerable popularity as a viable simulation framework for complex geometries compared to traditional mesh-based methods. Lagrangian mesh-free methods tend to operate at low-order convergence, whilst in an Eulerian frame, high-order approximations are possible, with the costs of calculating polynomially-consistent kernels incurred only in pre-processing.

In this work, we investigated the application of MLPs to surrogate segments of high-order mesh-free methods, specifically whether MLPs can obtain high-order weights with lower computational cost than current Eulerian mesh-free methods. In the first approach, we trained MLPs to predict the discretised weights in computational stencils from the distance of the support nodes to the central node in a computational stencil. The results of this approach indicate that, although MLPs can qualitatively learn the kernel present in high-order mesh-free methods with reasonable accuracy, the predicted weights fall short in eliminating undesired LABFM moments. Consequently, the ML-predicted outputs introduce amplified low-order errors, leading to relatively high limiting errors, only marginally outperforming uncorrected SPH. Hence, the architectures investigated are not suitable for high-order kernel surrogates. 

The second method proposed to accelerate LABFM involved utilising MLPs to solve the low-rank, dense, ill-conditioned and constant right-hand side linear system present in high-order mesh-free methods. Quantitatively, this approach yield significantly more accurate stencil moments than the first approach. The linear system surrogates generate significantly better results than the full discretisation surrogate, with stencil residuals about two orders of magnitude more accurate. The convergence profile of the trained models showed convergence at lower resolution before reaching a higher error limit than that obtained by LABFM. The limiting error of order $10^{-4}$ and $10^{-3}$ for the derivative and Laplacian, respectively, was attributed to the residual of the predicted solution of the linear system (i.e. stencil moments) compared to traditional methods, such as LU factorisation. The computational-cost accuracy trade-off study performed indicated that the linear system surrogate can be systematically optimised for accuracy or computational efficiency. In the case of computational efficiency optimisation, it was demonstrated that the linear system surrogate gives the same accuracy as traditional LU factorisation for a second-order approximation at lower computational cost up to the limiting accuracy. 

A current limitation is regarding the application of the current framework for high-order approximations. Given the condition of the linear systems deteriorates for higher-order approximations, the linear system must be solved with extremely high precision to achieve even moderate limiting accuracy. Thus, the framework proposed is currently not suitable for higher order approximations. Future directions that may circumvent these limitations include directly approximating the differential operators rather than the discrete weights, or embedding polynomial consistency as a hard constraints.

\section*{Data and Code Availability}

The data supporting the findings of this study will be submitted along with the manuscript. The GPU code used to train the neural network surrogate for the high-order kernel in mesh-free methods is available at \url{https://github.com/ML4MeshFree/train-hok.git}. The GPU code used to train neural networks to solve linear systems is available at \url{https://github.com/ML4MeshFree/train-ls.git}. The code used to surrogate the high-order kernel with neural networks is available at \url{https://github.com/ML4MeshFree/infer-hok.git}, and the code used to surrogate the linear system with neural networks is available at \url{https://github.com/ML4MeshFree/infer-ls.git}.

\section*{Declaration of Interest}
The authors declare that they have no known competing financial interests or personal relationships that could have appeared to influence the work reported in this paper.

\section*{Acknowledgements}
J. R. C. K is funded by the Royal Society via a University Research Fellowship (URF\textbackslash R1\textbackslash 221290). We would like to acknowledge the assistance given by Research IT and the use of the Computational Shared Facility at the University of Manchester. 

\renewcommand{\refname}{Bibliography}
\bibliographystyle{elsarticle-num-names} 
\bibliography{bibliography}

\appendix
\section{Smoothed Particle Hydrodynamics Operators}
\label{app:sph}
SPH gradient operators admit symmetric and anti-symmetric formulations, where the former guarantees exact conservation and the latter offers greater accuracy. Since the focus here is on consistency, only the anti-symmetric form is considered, which takes the form of Equation~\eqref{eq:discrete_op} with weights
\begin{equation}
\label{eq:sph_grad}
    w_{ji}^\nabla = \nabla_i W(\mathbf{x}_{ji},h)V_j,
\end{equation}
where $(\cdot)_{ji} = (\cdot)_j - (\cdot)_i$, so that $\mathbf{x}_{ji}$ denotes the position of node $j$ relative to node $i$, $W$ is the smoothing kernel, and $V_j$ is the particle volume. For the Laplacian, the Morris operator~\citep{Morris1997} is adopted:
\begin{equation}
    w_{ji}^\Delta = \frac{-2\,\mathbf{x}_{ji}}{||\mathbf{x}_{ji}||_2} \cdot \nabla_{i}W_{ji}\, V_{j},
\end{equation}
where $||\cdot||_2$ denotes the Euclidean norm.

\subsection{Smoothing Kernels}
\label{appendix:wendland}
\textbf{Quintic spline: } The quintic spline kernel is defined over the support $r \in [0,3]$ as
\begin{equation}
    W^\text{QS}(r) = \sigma
    \begin{cases}
        ( 1 - r)^5_+ - 6\left(\frac{2}{3} - r \right )^5_+ + 15 \left (\frac{1}{3} - r\right )^5_+, & 0 \le r < 1, \\
        ( 1 - r)^5_+ - 6\left(\frac{2}{3} - r \right )^5_+ , & 1 \le r < 2, \\
        ( 1 - r)^5_+, & 2 \le r < 3, \\
        0, & r \ge 3,
    \end{cases}
\end{equation}
where $r = ||\mathbf{x}_{ji}||_2/h$ and $\sigma$ is a normalisation constant dependent on spatial dimension; in two dimensions, $\sigma = \frac{7}{478\pi h^2}$~\citep{Liu2010SPH}. With $h = 1.5s$, the quintic spline yields approximately 60--65 neighbours.

\textbf{Wendland C2: } The Wendland C2 kernel is defined over the support $r \in [0,2]$ as
\begin{equation}
    W^{\text{WC}2}(r) = \sigma (1 - r)^4 (1 + 4r),
\end{equation}
with $\sigma = \frac{7}{\pi h^2}$ in two dimensions. With $h = 1.5s$, this kernel yields approximately 25--30 neighbours.

\section{Hyperparameter Details}
This appendix details the architectures and hyperparameters of all models used in this paper. All models were implemented using PyTorch's \texttt{nn.Module} class; any unspecified hyperparameter takes the PyTorch default value. Two distinct architectures were used, referred to as MLP and MLP$_s$, and are summarised in Table~\ref{table:hyp}. MLP$_s$ is used exclusively for the computational cost analysis in Section~\ref{sec:Linear sys w ML}; all other models use the MLP architecture.

\begin{table}[H]
\centering
\caption{Hyperparameters of models trained in this investigation}
\label{table:hyp}
\renewcommand{\arraystretch}{1.5}
\setlength{\tabcolsep}{12pt} 
\begin{tabular}{ccc}
  \toprule
    \textbf{Hyperparameter} & \textbf{$\text{MLP}$} & \textbf{$\text{MLP}_s$}  \\
    \toprule
    
  Hidden layers & 10 & 5 \\
    \hline
  Neurons/layer & 128 & 64  \\
  \hline
  Epochs & 2,000 & 1,500 \\
  \hline
  Batch size & 512 & 512  \\
  \hline
  Activation func. & SiLU & SiLU \\
  \hline
  Optimiser & Adam & Adam  \\
  \hline
  Loss func. & MSE & MSE  \\
  \hline
  Initial Learning rate & $10^{-3}$ & $10^{-3}$ \\
  \hline
  Optimiser Scheduler & Plateau Scheduler & Plateau Scheduler \\
  \bottomrule
\end{tabular}
\end{table}

\end{document}